# Compressive Sensing of Sparse Signals in the Hermite Transform Basis: Analysis and Algorithm for Signal Reconstruction

Miloš Brajović, *Student Member, IEEE*, Irena Orović, *Member, IEEE*, Miloš Daković, *Member, IEEE*
Srdjan Stanković, *Senior Member, IEEE*

*Abstract*—An analysis of the influence of missing samples in signals exhibiting sparsity in the Hermite transform domain is provided. Based on the statistical properties derived for the Hermite coefficients of randomly undersampled signal, the probability of success in detection of signal components support is determined. Based on the probabilistic analysis, a threshold for the detection of signal components is provided. It is a crucial step in the definition of a simple non-iterative algorithm for compressive sensing signal reconstruction. The derived theoretical concepts are proved on several examples using different statistical tests.

## I. INTRODUCTION

THE Hermite transform of signals has drawn significant research attention during the last decades, since it exhibits some important properties and high suitability for several signal processing applications [1]-[10]. Namely, the Hermite transform, referred also as the Hermite expansion, is an orthogonal signal representation with promising applicability in different research fields, due to its advantageous properties, such as the possibility of transform calculation via the recurrence relation as well as the property of the completeness of the Hermite basis. The Hermite functions have been recognized as a suitable basis for the representation and compression of QRS complexes in ECG signals [1]-[3]. Other important applications include: image processing, [4], [5], computed tomography, analysis of protein structure, optics [7], and radar signals [9]. Interesting mathematical properties of this transform have led to fast computation algorithms, which are important in state-of-the-art research in biomedicine and biology [1]. Their good localization properties have found important applications in time-frequency signal analysis, radar signal processing and processing of video signals [9],[10].

Also, previous studies addressed interesting mathematical issues such as the convergence properties and the optimum scaling of the Hermite expansion [7].

A small number of nonzero coefficients in a transform domain is the basic assumption for successful application of Compressive Sensing (CS) algorithms in the reconstruction of

The authors are with the Faculty of Electrical Engineering, University of Montenegro, 81000 Podgorica, Montenegro. The corresponding author is Irena Orović, phone:+38267516795, fax:+382245873 , Email: irenao@ac.me

signals with missing samples [11]-[28]. This useful property of a transform to represent the analyzed signals with small number of non-zero coefficients is identified as sparsity and measured by $\ell_0$-norm of the transform coefficients. When considering the Hermite transform, this assumption is valid for many of the mentioned types of signals, for instance the QRS complexes [1]-[3]. The reduced set of observations in CS is usually a consequence of sampling strategy, but signal samples can be intentionally omitted using robust signal processing due to high noise corruption [22],[23]. Therefore, our basic motivation is to analyze the influence of missing samples on the Hermite transform and signal reconstruction possibilities. The signal reconstruction is based on finding the solution of undetermined system of equations being the sparsest transform representation. Direct solution using minimization of $\ell_0$-norm is an NP-hard problem. In order to apply iterative minimization algorithms for finding the solution, or linear programming approaches and methods, the reconstruction constraint is relaxed, and $\ell_1$-norm is used as a measure of sparsity [12]-[14]. The solution can be obtained by using $\ell_1$-norm minimization via convex optimization algorithms, for example, primal-dual interior point methods. Other approaches are iterative procedures such as Orthogonal Matching Pursuit (OMP), Gradient Pursuit, CoSaMP [11]-[13], etc. An interesting iterative reconstruction algorithm which uses a steepest descent based procedure to achieve the minimization of the $\ell_1$-norm is used in [3]. Non-iterative approach for signal reconstruction that avoids the relaxation constraint is presented in [18]. It is based on the comprehensive analysis of the missing samples influence to the sparse transform, namely, the Discrete Fourier Transform [22]. However, due to the specific form and different properties of the Hermite transform, direct generalization of the mentioned reconstruction approach to this sparsity domain is not possible. This fact led us to the theoretical contributions presented in this study.

The paper is organized as follows: the Hermite transform and its placement into the CS framework is done in Section II. Detailed analysis of the missing samples influence on the Hermite transform is provided in Section III. Theory extension towards the simple reconstruction approach is done in Section IV. Section V provides numerical evaluation of the presented theory along with reconstruction example, while the concluding remarks are given in Section VI.

## II. THEORETICAL BACKGROUND

### A. Discrete Hermite transform

Hermite polynomial of the *p*-th order, widely known among the orthogonal polynomials, can be defined as [1]-[8]:

$$H_p(t) = (-1)^p e^{t^2} \frac{d^p(e^{-t^2})}{dt^p}. \quad (1)$$

The *p*-th order Hermite function is related with the *p*-th order Hermite polynomial as follows:

$$\psi_p(t,\sigma) = \left(\sigma 2^n n! \sqrt{\pi}\right)^{-1/2} e^{-t^2/2} H_p(t/\sigma), \quad (2)$$

where the constant $\sigma$ is used to "stretch" and "compress" Hermite functions, in order to provide a representation with desirable properties [1]. In further analysis, for the sake of simplicity, it will be assumed that this constant is $\sigma = 1$. The Hermite functions can be calculated in a recursive manner, which is an advantage in applications [1], [6]. The orthogonality of the Hermite polynomials and the orthonormality of Hermite functions, often makes them suitable as a basis for signal representation. The Hermite expansion or Hermite transform is given by [1]-[6]:

$$f(t) = \sum_{p=0}^{N} c_p \psi_p(t,\sigma), \quad (3)$$

where $c_p$ denotes the *p*-th order Hermite coefficient:

$$c_p = \int_{-\infty}^{\infty} f(t) \psi_p(t) dt, \; p = 0,1,...,M-1. \quad (4)$$

An infinite number $N \to \infty$ of Hermite functions is needed for the exact representation of the continuous signal *f(t)*. However, in numerous applications, a finite number of *N* Hermite functions can be used with a certain approximation error, e.g. [1], [2], [10]. For the numerical calculation of the integral (4) quadrature approximation techniques have been used, [1], [8], [9] and usually interpreted as discrete form of the Hermite transform. Since it provides significant calculation advantages over other approximations, the Gauss-Hermite quadrature can be considered:

$$c_p = \frac{1}{M} \sum_{m=1}^{M} \frac{\psi_p(t_m)}{[\psi_{M-1}(t_m)]^2} f(t_m), \; p = 0,1,...,M-1, \quad (5)$$

where $t_m$ is used to denote zeros of the *M*-th order Hermite polynomial. If continuous Hermite functions are sampled at the zeros of the *M*-th order Hermite polynomial, then the summation (3) becomes a finite orthonormal representation of the analyzed signal. For a signal of length *M*, the complete set of discrete Hermite functions used for unique signal representation consists of exactly *M* functions [1], [3]. In certain applications, such as image processing, a smaller number of Hermite functions $N < M$ can be used [14].

Note that the discrete Hermite transform satisfies the orthonormality property [1], [3], [7]:

$$\frac{1}{M} \sum_{m=1}^{M} \frac{\psi_p(t_m)}{(\psi_{M-1}(t_m))^2} \psi_k(t_m) = \delta(p-k), \quad (6)$$

and:

$$\frac{1}{M} \sum_{p=0}^{M-1} \frac{\psi_p(t_m)}{(\psi_{M-1}(t_m))^2} \psi_p(t_n) = \delta(n-m) \quad (7)$$

with *m* and *n* being the indices of the Hermite polynomial roots $t_m$ and $t_n$ respectively. In further analysis, it will be assumed that the analyzed signal (of length *M*) and Hermite functions are sampled at Hermite polynomial roots $t_m$ and the index *m* will be used to denote the discrete time index.

Having in mind the previous analysis, the expansion using *M* Hermite functions can be written in matrix-vector notation. Let us introduce the vector $\mathbf{c} = [c_0, c_1, ..., c_{M-1}]^T$ consisted of Hermite coefficients $c_p$, and vector $\mathbf{f} = [f(1), f(2), ..., f(M)]^T$ consisted of *M* signal samples. Having in mind the Gauss-Hermite approximation formula (5), the inverse transform matrix $\mathbf{\Psi}$ is consisted of *M* Hermite functions:

$$\mathbf{\Psi} = \begin{bmatrix} \psi_0(1) & \psi_0(2) & \cdots & \psi_0(M) \\ \psi_1(1) & \psi_1(2) & \cdots & \psi_1(M) \\ \vdots & \vdots & \ddots & \vdots \\ \psi_{M-1}(1) & \psi_{M-1}(2) & \cdots & \psi_{M-1}(M) \end{bmatrix}.$$

Based on previous matrix definition, the Hermite transform for the case of discrete signals can be written as:

$$\mathbf{f} = \mathbf{\Psi} \mathbf{c}. \quad (8)$$

### B. Compressive sensing and Hermite transform

The compressive sensing procedure based on the random selection/acquisition of signal values can be modeled by using a random measurement matrix $\mathbf{\Phi}$:

$$\mathbf{y}_{cs} = \mathbf{\Phi} \mathbf{f} = \mathbf{\Phi} \mathbf{\Psi} \mathbf{c} = \mathbf{A}_{cs} \mathbf{c},$$

where $\mathbf{y}_{cs}$ denotes the vector of available samples of the analyzed signal. The matrix $\mathbf{A}_{cs}$ is obtained from the inverse transform matrix $\mathbf{\Psi}$, in our case the inverse Hermite transform matrix, by omitting the rows corresponding to the positions of missing samples. The available samples have random positions denoted by:

$$m \in \mathbf{M}_A = \{m_1, m_2, ..., m_{M_A}\} \subseteq \mathbf{M} = \{1, 2, ..., M\}. \quad (9)$$

Note again that the index *m* on the discrete grid corresponds to the sampling point $t_m$. In order to obtain the reconstructed signal values, an undetermined system of $M_A$ linear equations and *M* unknowns have to be solved. It is known that such systems may have infinitely many solutions, but the idea behind the compressive sensing is to find the sparsest one. The signal reconstruction problem is usually reduced to the problem of identifying signal support (positions and values of non-zero coefficients in the sparsity domain). Here, we assume that the observed signal is sparse in the Hermite transform domain, i.e., $K \ll M$ and *K* being the number of nonzero Hermite coefficients. The non-zero coefficients have indices from the set:

$$\mathbf{P} = \{p_1, p_2, ..., p_K\} \subseteq \mathbf{P}_M = \{0, 1, ..., M-1\}, K \ll M.$$

Finding the sparsest solution corresponds to solving the

minimization of the form

$$\min \|\mathbf{c}\|_0 \text{ subject to } \mathbf{y}_{cs} = \mathbf{A}_{cs}\mathbf{c}. \quad (10)$$

It is known that the $\ell_0$-norm cannot be used in the direct minimization and thus the problem (10) is usually reformulated using $\ell_1$-norm whose convexity enables application of efficient linear programming and iterative approaches. On the other side, if the signal support is known or appropriately estimated within a set $\hat{\mathbf{P}}$ containing $K \leq \hat{K} \leq M$ elements such that $\mathbf{P} \subseteq \hat{\mathbf{P}}$, the reconstruction is achieved using the pseudo-inversion:

$$\mathbf{c}_K = \left(\mathbf{A}_{csK}^T \mathbf{A}_{csK}\right)^{-1} \mathbf{A}_{csK}^T \mathbf{y}_{cs}. \quad (11)$$

The matrix $\mathbf{A}_{csK}$ is the sub-matrix of the matrix $\mathbf{A}_{cs}$ with omitted columns corresponding to positions $p \notin \hat{\mathbf{P}}$.

Our aim is to analyze the influence of missing samples of the compressed sensed signal to the Hermite domain representation. If we are able to model and characterize the effects caused in the sparsity domain as a consequence of compressive sampling, then we can develop an efficient procedure to determine signal support in the transform domain, defined by a proper set $\hat{\mathbf{P}}$ suitable for the reconstruction.

### III. ANALYSIS OF MISSING SAMPLES

Consider the Hermite transform of the signal $s(m)$ sampled at the points corresponding to the zeros of the $M$-th order Hermite polynomial. The coefficients of the Hermite expansion using $M$ Hermite functions are calculated by:

$$c_p = \frac{1}{M}\sum_{m=1}^{M} \mu_{M-1}^p(m)s(m) = \frac{1}{M}\sum_{m=1}^{M} \frac{\psi_p(m)}{(\psi_{M-1}(m))^2} s(m) \quad (12)$$

The fact that the signal samples are placed on a grid corresponding to Hermite polynomials zero allows a high level accuracy in the Gauss-Hermite quadrature calculation. We will assume that the analyzed signal $s(m)$ is sparse in the Hermite domain so that it can be represented as:

$$s(m) = \sum_{i=1}^{K} A_i \psi_{p_i}(m) \quad (13)$$

with $K$ being the number of signal components, $A_i$ is used to denote amplitudes of signal component, $p_i$ denotes the order of the Hermite function. For the multicomponent signal (13) the Hermite transform coefficients (12) are calculated as follows:

$$c_p = \sum_{m=1}^{M}\sum_{i=1}^{K} \frac{A_i}{M} \frac{\psi_p(m)\psi_{p_i}(m)}{(\psi_{M-1}(m))^2}, p = 0,...,M-1. \quad (14)$$

Normalized signal components are multiplied by the orthonormal basis functions $\mu_{M-1}^p(m) = \psi_p(m)/(\psi_{M-1}(m))^2$ to produce the signal $y_p(m)$ defined as:

$$y_p(m) = \frac{A_i}{M} \frac{\psi_p(m)\psi_{p_i}(m)}{(\psi_{M-1}(m))^2}. \quad (15)$$

Values of the signal denoted with $y_p(m)$ are from the set:

$$\mathbf{\Omega} = \left\{ \frac{A_i}{M} \frac{\psi_p(m)\psi_{p_i}(m)}{(\psi_{M-1}(m))^2}, m=1,...,M \right\}. \quad (16)$$

Since the orthonormality property (6) holds, it is obvious that the members of $\mathbf{\Omega}$ satisfy the relation:

$$\sum_{m=1}^{M} y_p(m) = y_p(1) + y_p(2) + ... + y_p(M) = 0 \quad (17)$$

for given $p \neq p_i$, $i = 1, 2..., K$.

In order to analyze the CS signal case, a subset consisted of $M_A \leq M$ randomly positioned available samples from the set $\mathbf{\Omega}$ is considered:

$$\mathbf{\Theta} = \{y_p(m_1), y_p(m_2), ..., y_p(m_{M_A})\} \subseteq \mathbf{\Omega}. \quad (18)$$

Thus, $M_Q = M - M_A$ samples are unavailable. Since the Hermite transform is a linear operator, and the inner products are performed between signal values and the basis functions, if some samples are omitted from the signal, it produces the same result as if these samples assume zero values. Consequently, a reduced number of signal samples can be considered as a complete set of samples, where some of them are affected by noise modeled as:

$$\eta(m) = \begin{cases} -y_p(m), & \text{for } m \in \mathbf{N} \setminus \mathbf{M}_A \\ 0, & \text{for } m \in \mathbf{M}_A \end{cases}$$

Under this assumption, in the sequel we will derive the statistical properties of the Hermite transform coefficients on the signal and non-signal positions.

*A. Mono-component signal case*

First the one-component signal case, with $K=1$, $A_i=1$ and $p_i = p_0$, will be considered. The Hermite transform over the set of available samples from $\mathbf{\Theta}$ can be written in the following form:

$$Y_p = c_p \approx \sum_{i=1}^{M_A} y_p(m_i) = \sum_{m=1}^{M} [y_p(m) + \eta(m)]. \quad (19)$$

It is a random variable, formed as a sum of $M_A$ randomly positioned available samples. Here the derivation of the mean value and the variance of the random variable $Y_p$ in the Hermite domain will be conducted. As it will be shown, this variable has different statistical properties at the position $p=p_0$ corresponding to the signal component, and at other positions $p \neq p_0$ in the Hermite domain corresponding to the noise.

*1) Statistical properties of the Hermite transform at the non-signal positions*

For the non-signal positions $p \neq p_0$, the random variable $Y_{p \neq p_0}$ corresponds to an additive transform domain noise [22]. Having in mind the orthonormality property (6) and the fact that samples $y_p(m_i)$ from the set $\mathbf{\Theta}$ have random positions, it is obvious that $E\{y_p(m_i)\} = E\{y_p(m)\} = 0$ with a high probability. Thus, the random variable $Y_{p \neq p_0}$ has a zero mean value:

$$\mu_N = E\{Y_{p \neq p_0}\} = E\left\{\sum_{i=1}^{M_A} y_p(m_i)\right\} = \sum_{i=1}^{M_A} E\{y_p(m_i)\} = 0. \quad (20)$$

The variance of $Y_{p \neq p_0}$, taking into account that it is real-valued is defined as follows:

$$\sigma_N^2 = \text{var}\{Y_{p \neq p_0}\} = E\{|Y_{p \neq p_0}|^2\} = E\{(Y_{p \neq p_0})(Y_{p \neq p_0})^*\} =$$
$$= E\left\{\sum_{i=1}^{M_A}\sum_{j=1}^{M_A} y_p(m_i) y_p(m_j)\right\} = \quad (21)$$
$$= \underbrace{E\left\{\sum_{i=1}^{M_A} y_p^2(m_i)\right\}}_{S1} + \underbrace{E\left\{\sum_{i=1}^{M_A}\sum_{\substack{j=1 \\ i\neq j}}^{M_A} y_p(m_i) y_p(m_j)\right\}}_{S2}.$$

According to (17), we have:
$$E\{y_p(m_i)(y_p(1) + y_p(2) + \ldots + y_p(M))\} = 0 \quad (22)$$
i.e., $E\{y_p(m_i)y_p(1)\} + E\{y_p(m_i)y_p(2)\} + \ldots + E\{y_p(m_i)y_p(M)\} = 0$, for $i = 1, 2, \ldots, M$.

The terms $E\{y_p(m_i)y_p(m_j)\}$, for $i \neq j$ are equally distributed:
$$E\{y_p(m_i)y_p(m_j)\} = B, \quad i \neq j. \quad (23)$$

The expectation $E\{y_p(m_i)y_p(m_j)\}$ for $i \neq j$ can be derived from the orthonormality property (7), and the fact that $p \neq p_0$ is considered:
$$E\{y_p(m_i)y_p(m_i)\} = E\left\{\frac{1}{M^2}\frac{\psi_p(m_i)\psi_p(m_i)}{(\psi_{M-1}(m_i))^2}\frac{\psi_{p_0}(m_i)\psi_{p_0}(m_i)}{(\psi_{M-1}(m_i))^2}\right\} =$$
$$= E\left\{\frac{1}{M}\frac{\psi_p(m_i)\psi_p(m_i)}{(\psi_{M-1}(m_i))^2}\right\} E\left\{\frac{1}{M}\frac{\psi_{p_0}(m_i)\psi_{p_0}(m_i)}{(\psi_{M-1}(m_i))^2}\right\}.$$

Note that the appearances of the Hermite functions of orders $p$ and $p_0$ (i.e. their values at instant $m_i$) are statistically independent events, and thus the expectations can be separated. Due to orthogonality property (7) we have:
$$E\left\{\frac{1}{M}\frac{\psi_p(m_i)\psi_p(m_i)}{(\psi_{M-1}(m_i))^2}\right\} = E\left\{\frac{1}{M}\frac{\psi_{p_0}(m_i)\psi_{p_0}(m_i)}{(\psi_{M-1}(m_i))^2}\right\} = \frac{1}{M}. \quad (24)$$

Thus it can be easily concluded that:
$$E\{y_p(m_i)y_p(m_i)\} = E\{y_p(m_i)\}E\{y_p(m_i)\} = 1/M^2. \quad (25)$$

Since there are $M-1$ terms given by (23) with the same expectation $B$ for $i \neq j$ and one term with value (25) for $i = j$, from (22) follows: $1/M^2 + (M-1)B = 0$.

The unknown $B$ is therefore given by:
$$E\{y_p(m_i)y_p(m_j)\} = B = \frac{-1}{M^2(M-1)}, \quad i \neq j \quad (26)$$

Since there are $M_A$ terms in $S1$ summation in (21) and $M_A(M_A - 1)$ terms in $S2$ summation, we finally obtain the variance at the non-signal positions (noise variance) as:
$$\sigma_N^2 = \text{var}\{Y_{p \neq p_0}\} = \frac{M_A}{M^2} + M_A(M_A - 1)B = \frac{M_A M - M_A^2}{M^2(M-1)} \quad (27)$$

We can conclude that the variance of the noise at non-signal positions $p \neq p_0$ in the Hermite domain depends only on the number of available samples $M_A$ and the signal length $M$. According to the central limit theorem, the observed random variable $Y_{p \neq p_0}$ has the normal distribution.

*2) Statistical properties of Hermite transform at the signal components positions*

The statistics of the Hermite expansion coefficients of the CS signal for the case $p = p_0$ is quite different. Since that the product $\psi_{p_0}(m_i)\psi_{p_0}(m_i)$ in the considered case depends on the values of the specific Hermite function $\psi_{p_0}(m_i)$, whose samples are missing at random positions, it is obvious that $Y_{p=p_0}$ is also a random variable with normal distribution, according to the central limit theorem.

In the case of $y_{p=p_0}(m_i) = y_{p_0}(m_i)$ with $M_A$ available randomly positioned samples from the set $\Theta$, the expected (mean) value of the random variable $Y_{p=p_0}$ follows from (24):
$$\mu_s = E\{Y_{p=p_0}\} = \frac{1}{M}E\left\{\sum_{i=1}^{M_A}\frac{\psi_{p_0}(m_i)\psi_{p_0}(m_i)}{(\psi_{M-1}(m_i))^2}\right\} = \frac{M_A}{M} \quad (28)$$

and all the values of the random variable $Y_{p=p_0}$ are equally distributed. Since the mean value is not equal to zero, variance is calculated as follows:
$$\sigma_s^2 = \text{var}\{Y_{p=p_0}\} = E\{|Y_{p=p_0} - \mu_s|^2\} = E\{Y_{p=p_0}^2\} - |\mu_s|^2. \quad (29)$$

Using the definition of the random variable $Y_{p=p_0}$, the variance $\sigma_s^2$ can be expanded in the form:
$$\sigma_s^2 = E\left\{\sum_{i=1}^{M_A} y_{p_0}^2(m_i)\right\} + E\left\{\sum_{i=1}^{M_A}\sum_{\substack{j=1 \\ i\neq j}}^{M_A} y_{p_0}(m_i)y_{p_0}(m_j)\right\} - |\mu_s|^2 \quad (30)$$

The calculation of individual terms in (30) will differ from the previous case ($p \neq p_0$). Starting from the orthogonality property (7) for $p = p_0$:
$$y_{p_0}(1) + y_{p_0}(2) + \ldots + y_{p_0}(M) = 1 \quad (31)$$

and multiplying left and right side by $y_{p_0}(m_i), i \in \{1,\ldots,M\}$, the expectation is calculated as:
$$E\{y_{p_0}(m_i)y_{p_0}(1) + y_{p_0}(m_i)y_{p_0}(2) + \ldots + y_{p_0}(m_i)y_{p_0}(M)\} = E\{y_{p_0}(m_i)\} \quad (32)$$
i.e., $E\{y_{p_0}(m_i)y_{p_0}(1)\} + \ldots + E\{y_{p_0}(m_i)y_{p_0}(M)\} = 1/M \quad (33)$

In the case $i \neq j$, $M - 1$ terms are equally distributed and satisfy:
$$E\{y_{p_0}(m_i)y_{p_0}(1)\} = \ldots = E\{y_{p_0}(m_i)y_{p_0}(M)\} =$$
$$= E\{y_{p_0}(m_i)y_{p_0}(m_j)\} = D$$

For $i = j$ the expectations $E\{y_{p_0}^2(m_i)\}$, $i = 1,\ldots M$ cannot be estimated as $E\{y_{p_0}(m_i)\}E\{y_{p_0}(m_i)\}$, because statistical independence requirement is not satisfied. Hence, in order to determine $E\{y_{p_0}^2(m_i)\}$ let us observe the summation:
$$E\left\{\sum_{i=1}^{M} y_{p_0}^2(m_i)\right\} = \frac{1}{M^2}\sum_{i=1}^{M}\left(\frac{\psi_{p_0}^2(m_i)}{(\psi_{M-1}(m_i))^2}\right)^2 =$$
$$= \frac{1}{M^2}\sum_{i=1}^{M} a(m_i, p_0, M) = \frac{1}{M^2} P_{p_0} \quad (34)$$

which corresponds to the energy of the mono-component signal defined by the Hermite function of order $p_0$. Note that the following notation is used:

$$\left(\frac{\psi_{p_0}^2(m_i)}{(\psi_{M-1}(m_i))^2}\right)^2 = a(m_i, p_0, M)$$

where $m_i \in \{m_1, m_2, ..., m_{M_A}\}$ are random positions of $M_A$ available samples. It can be concluded that:

$$E\{a(m_i, p_0, M)\} = E\{a(p_0, M)\} = P_{p_0}/M^3.$$

Now, (33) becomes:

$$a(p_0, M) + (M-1)D = 1/M,$$

and then $D$ can be expressed as:

$$D = \frac{1 - Ma(p_0, M)}{M(M-1)}. \quad (35)$$

The variance (30) of the considered random variable can be now written as:

$$\sigma_s^2 = M_A a(p_0, M) + M_A(M_A - 1)D - \left(\frac{M_A}{M}\right)^2 = $$
$$= M_A a(p_0, M) + M_A(M_A - 1)\frac{1 - Ma(p_0, M)}{M(M-1)} - \left(\frac{M_A}{M}\right)^2 \quad (36)$$

After simple rearrangement of the previous equation, the variance can be expressed as:

$$\sigma_s^2 = \text{var}\{Y_{p=p_0}\} = \frac{MM_A - M_A^2}{M^2(M-1)}(M^2 a(p_0, M) - 1)$$
$$= \sigma_N^2 \left(\frac{1}{M}\sum_{m_i=1}^{M}\left(\frac{\psi_{p_0}^2(m_i)}{(\psi_{M-1}(m_i))^2}\right)^2 - 1\right) = \sigma_N^2\left(\frac{P_{p_0}}{M} - 1\right) \quad (37)$$

The relation (37) describing how the variance depends on the Hermite coefficient order $p_0$ is evaluated also experimentally. The results are shown in Fig. 1, for the signal of length $M = 200$, with $M_A = 120$ available samples. The numerical calculation of the variance is obtained using 5000 independent realizations of the signal with randomly positioned missing samples.

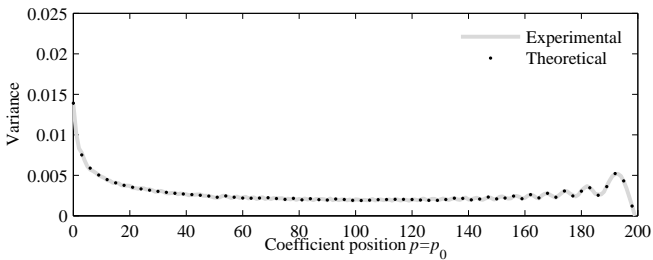

Fig. 1 The variance at the position of the signal component as a function of the component position $p_0$

When only $M_A$ out of $M$ samples are available, the known bias in the amplitude should be compensated by $M/M_A$, while $P_{p_0}$ can be estimated from the available set of samples:

$$\tilde{P}_{p_0} = \sum_{i=1}^{M_A}\left(\frac{\psi_{p_0}^2(m_i)}{(\psi_{M-1}(m_i))^2}\right)^2.$$

Consequently, the variance at the signal component positions $p=p_0$, for an incomplete set of samples, can be estimated as:

$$\tilde{\sigma}_s^2 = \frac{M}{M_A}\sigma_N^2\left(\frac{\tilde{P}_{p_0}}{M} - 1\right). \quad (38)$$

*B. Probabilistic analysis of detection error for Hermite coefficient corresponding to signal component*

According to the central limit theorem, both random variables $Y_{p=p_0}$ and $Y_{p\neq p_0}$ behave as Gaussian variables with their own mean values and variances. The derived mean values and variances will be used to define a method to distinguish between Hermite transform components corresponding to signal from those corresponding to noise caused by missing samples. This approach refers to the signal component detection. In the sequel, we consider the absolute values of the random variables $Y_{p=p_0}$ and $Y_{p\neq p_0}$. Given a normally distributed random variable $Y_{p=p_0}$ corresponding to the signal component in the Hermite transform domain, with mean value $\mu_s$ and variance $\sigma_s^2$ (given by (28) and (38) respectively), the random variable $\xi = |Y_{p=p_0}|$ has the Folded Normal Distribution as the probability density function (pdf):

$$f(\xi) = \frac{1}{\sigma_s\sqrt{2\pi}}\left(\exp\left(-\frac{(\xi - \mu_s)^2}{2\sigma_s^2}\right) + \exp\left(-\frac{(\xi + \mu_s)^2}{2\sigma_s^2}\right)\right), \quad (39)$$

see Fig. 2a. The random variable which corresponds to the noise, $Y_{p\neq p_0}$, has also the normal pdf, while its absolute value, $\zeta = |Y_{p\neq p_0}|$ has the Half Normal Distribution, since the mean value is zero:

$$\eta(\zeta) = \frac{\sqrt{2}}{\sigma_N\sqrt{\pi}}\exp\left(-\frac{\zeta^2}{2\sigma_N^2}\right), \quad (40)$$

with variance given by (27). This distribution along with the experimentally obtained histogram is shown in Fig. 2b.

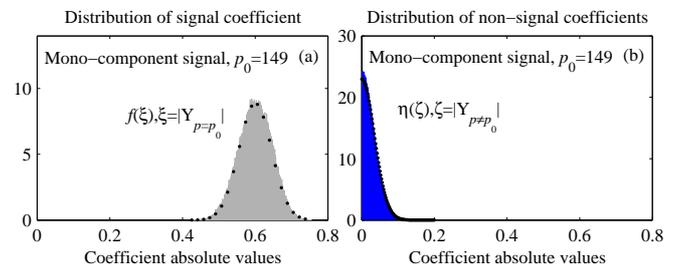

Fig. 2: Histograms and pdfs for the absolute values of Hermite coefficients at: (a) signal and (b) non-signal positions. Histograms are simulated for signal with $M_A=120$ out of $M=200$ samples and amplitude $A_0=1$, based on 20000 independent signal realizations with randomly positioned missing samples. Theoretical results (dots) are obtained using Folded Normal Distribution (39) calculated with estimated value of variance (38) and Half Normal Distribution (40) with variance (27).

The probability that the random variable $\zeta = |Y_{p \neq p_0}|$ is smaller than $\chi$ is:

$$P_N(\chi) = \int_0^\chi \frac{\sqrt{2}}{\sigma_N \sqrt{\pi}} \exp\left(-\frac{\zeta^2}{2\sigma_N^2}\right) d\zeta = \mathrm{erf}\left(\frac{\chi}{\sqrt{2}\sigma_N}\right). \quad (41)$$

The total number of noise-alone components is $M-1$. Probability that $M-1$ independent noise components are smaller than $\chi$ is:

$$P_{NN}(\chi) = \mathrm{erf}\left(\frac{\chi}{\sqrt{2}\sigma_N}\right)^{M-1}. \quad (42)$$

The probability that at least one noise component is larger than $\chi$ is $P_{NL}(\chi) = 1 - P_{NN}(\chi)$.

If the signal value is within $\xi$ and $\xi + d\xi$ with probability $f(\xi)d\xi$, it will be misdetected if at least one noise component is above $\xi$. This event will occur with the probability $P_{NL}(\xi)f(\xi)d\xi = (1-P_{NN}(\xi))f(\xi)d\xi$. Considering all possible values of $\xi$, the misdetection will occur with probability:

$$P_E = \int_0^\infty (1 - P_{NN}(\xi)) f(\xi) d\xi = \frac{1}{\sigma_s \sqrt{2\pi}} \int_0^\infty \left(1 - \mathrm{erf}\left(\frac{\xi}{\sqrt{2}\sigma_N}\right)^{M-1}\right)$$
$$\times \left(\exp\left(-\frac{(\xi - \mu_s)^2}{2\sigma_s^2}\right) + \exp\left(-\frac{(\xi + \mu_s)^2}{2\sigma_s^2}\right)\right) d\xi \quad (43)$$

Previous relation is the probability of error in the detection of signal component (misdetection) for a one component sparse signal. It can be approximated using the assumption that the signal component is deterministic, and equal to its mean value $\mu_s = E\{Z_{p=p_0}\}$. A rough approximation of the error probability follows [1]:

$$P_E \approx 1 - \mathrm{erf}\left(\frac{\mu_s}{\sqrt{2}\sigma_N}\right)^{M-1}. \quad (44)$$

This approximation can be corrected with 1.5 standard deviation of the signal component if we use the fact that signal components in Hermite domain smaller than the mean value contribute more to the error than those above the mean value:

$$P_E \approx 1 - \mathrm{erf}\left(\frac{\mu_s - 1.5\sigma_s}{\sqrt{2}\sigma_N}\right)^{M-1}. \quad (45)$$

Note that in the case of signal with non-unit amplitude $A_0$, mean value is multiplied by the amplitude, while the signal variance is multiplied by $A_0^2$, for both analyzed cases.

---

[1] The expected value and variance of a random variable with Folded Normal Distribution are given as:

$$E\{\zeta\} = \sigma_s \sqrt{\frac{2}{\pi}} \exp\left(\frac{-\mu_s^2}{2\sigma_s^2}\right) - \mu_s \mathrm{erf}\left(\frac{-\mu_s}{\sqrt{2}\sigma_s}\right) \cong \mu_s.$$

$$\mathrm{var}\{\zeta\} = \mu_s^2 + \sigma_s^2 - \left\{\sigma_s \sqrt{\frac{2}{\pi}} \exp\left(\frac{-\mu_s^2}{2\sigma_s^2}\right) - \mu_s \mathrm{erf}\left(\frac{-\mu_s}{\sqrt{2}\sigma_s}\right)\right\} \cong \sigma_s^2.$$

### C. Analysis of multicomponent signals

The previous analysis will be extended to the multicomponent signals. In the case of multicomponent signals, a new random variable $y_p(m) \in \Omega$ can be introduced:

$$y_p(m) = \sum_{l=1}^K \frac{A_l}{M} \frac{\psi_p(x_m)\psi_{p_l}(x_m)}{(\psi_{M-1}(x_m))^2} \quad (46)$$

which consists of $K$ components. According to the previous results, in the case of $K$-component signal, the value of the coefficients at the signal position, $Y_{p=p_i}$ behave as a Gaussian variable, with mean value equal to:

$$\mu_S = \sum_{l=1}^K A_l \frac{M_A}{M} \delta(p - p_l), \quad (47)$$

since the noise caused by missing samples is zero mean, as shown for the mono-component case.

The variance at the points with no signal components is equal to:

$$\sigma_N^2 = \mathrm{var}\{Y_{p \neq p_i}\} = \sum_{l=1}^K \sigma_{N,l}^2 = \frac{M_A M - M_A^2}{M^2(M-1)} \sum_{l=1}^K A_l^2, \quad (48)$$

since at the points $p \neq p_i$ the noise caused by missing samples from each signal component contributes, and these noisy components are uncorrelated and zero mean.

According to the presented mono-component analysis, the $i$-th signal component at the position $p = p_i$ has the variance equal to:

$$\tilde{\sigma}_{s_i}^2 \approx \frac{M_A M - M_A^2}{M^2(M-1)} \frac{M}{M_A} A_i^2 \left(\sum_{i=1}^{M_A} \left(\frac{\psi_{p_i}^2(m_i)}{(\psi_{M-1}(m_i))^2}\right)^2 - 1\right)$$

where the subscript $s_i$ denotes that it originates from the $i$-th signal component. Additionally, the noise caused by missing samples from other $K-1$ components is also present at the position of the $i$-th signal component. This means that, besides the random variable $Y_{p=p_i}$, the sum of $K-1$ random variables $Y_{p \neq p_i}$ with $p \in \mathbf{P} = \{p_1, p_2, ..., p_K\}$ originating from other signal components also affects the $i$-th signal position. Note that all random variables at the $i$-th position are normally distributed. $K-1$ random variables $Y_{p \neq p_i}, p \in \{p_1, p_2, ..., p_K\}$ are zero mean, while the random variable $Y_{p=p_i}$ has the mean value $\mu_{S,i} = A_i M_A / M$. The resulting variance at the $i$-th signal position $p=p_i$ is finally:

$$\sigma_i^2 = \tilde{\sigma}_{s_i}^2 + \sum_{\substack{l=1 \\ l \neq i}}^K \sigma_{N,l}^2 = \frac{M - M_A}{M(M-1)}\left(A_i^2\left(\frac{\tilde{P}_{p_i}}{M} - 1\right) + \sum_{\substack{l=1 \\ l \neq i}}^K A_l^2\right). \quad (49)$$

It can be concluded that the Hermite expansion coefficient at signal position $p=p_i$ will be modelled as the random variable $Y_{p=p_i}$ with normal distribution: $\mathcal{N}(M_A A_i / M, \sigma_i^2)$. Also, the coefficients corresponding to noise are modelled by the random variable $Y_{p \neq p_i}$ with normal distribution: $\mathcal{N}(0, \sigma_N^2)$ where $\sigma_N^2$ is given by (48).

As it is done for the mono-component signal case, previous results can be used to derive the probability of error in the detection of signal components. The false signal component detection occurs when at least one noise component at positions $p \neq p_i, i \in \{1, 2, ..., K\}$ is above signal component at the position $p_i$. Recall that the absolute values of the random variables $Y_{p=p_i}$ and $Y_{p \neq p_i}$ have Half Normal and Folded Normal distributions, as shown on Fig. 3a and Fig. 3b respectively.

The random variable $\xi = |Y_{p=p_i}|$ has the pdf given by (39), with mean value equal to (47) and variance given by (49). The random variable representing the noise $\zeta = |Y_{p \neq p_0}|$, is zero mean, with Half Normal pdf (40), where the variance is equal to (48). The probability that $M - K$ independent noise alone points are smaller than $\chi$ is:

$$P_{NN}(\chi) = \text{erf}\left(\frac{\chi}{\sqrt{2}\sigma_N}\right)^{M-K}. \quad (50)$$

Following the mono-component signal case, the probability of error in the detection of the $i$-th signal component in the multicomponent case has the form:

$$P_{E_i} = \frac{1}{\sigma_i\sqrt{2\pi}} \int_0^\infty \left(1 - \text{erf}\left(\frac{\xi}{\sqrt{2}\sigma_N}\right)^{M-K}\right) \\ \times \left(\exp\left(-\frac{(\xi - \mu_{s_i})^2}{2\sigma_i^2}\right) + \exp\left(-\frac{(\xi + \mu_{s_i})^2}{2\sigma_i^2}\right)\right) d\xi \quad (51)$$

Under the same assumptions as in the mono-component signal case, this error can be approximated by:

$$P_{E_i} \approx 1 - \text{erf}\left(\frac{\mu_{s_i} - 1.5\sigma_i}{\sqrt{2}\sigma_N}\right)^{M-K}. \quad (52)$$

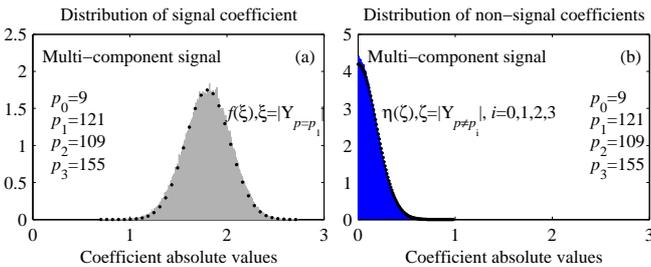

Fig. 3: Histograms and pdfs for the absolute values of Hermite coefficients at: (a) signal and (b) non-signal positions. Histograms are simulated for multicomponent signals with $M_A=120$ out of $M=200$ samples and amplitudes $A_0=1$, $A_1=3$, $A_2=4$ and $A_3=2$, based on 20000 independent signal realizations with randomly positioned available samples. Theoretical results are obtained using Folded Normal Distribution calculated with estimated value of variance (49), and Half Normal Distribution with variance (48).

## IV. DETECTION OF SIGNAL COMPONENTS AND SIGNAL RECONSTRUCTION ALGORITHM

### A. Detection of signal component in the Hermite transform domain

Due to its importance, we will consider in detail the probability that $M - K$ independent noise components are smaller than $\chi$ given by (50). This relation can give the threshold $\chi=T$ for the separation of signal components and noise. Following (50) for $\chi=T$, the threshold value can be derived as follows:

$$T = \sqrt{2}\sigma_N \text{erf}^{-1}\left(\left(P_{NN}(T)\right)^{\frac{1}{M-K}}\right) \approx \sqrt{2}\sigma_N \text{erf}^{-1}\left(\left(P_{NN}(T)\right)^{\frac{1}{M}}\right). (53)$$

Note that $K$ can be neglected in (53), since the number of components $K$ is in general much lower than the number of samples $M$ ($K<<M$). The threshold is calculated for a given (desired) probability $P_{NN}(T)$, using the noise variance defined by (48). Furthermore, the function $\text{erf}(x)$ can be approximated by:

$$\text{erf}(x) \approx \text{sgn}(x)\sqrt{1 - \exp\left(-x^2 \frac{4/\pi + ax^2}{1 + ax^2}\right)}, \quad (54)$$

with $a \approx 0.147$, and $x = T/(\sqrt{2}\sigma_N)$. Since $T \geq 0$ and $\sigma_N \geq 0$, and thus $x>0$ we conclude that it always holds that $\text{sgn}(x)=1$. Then, according to (54), we have:

$$\left(P_{NN}(T)\right)^{\frac{1}{M}} = \sqrt{1 - \exp\left(-x^2 \frac{4/\pi + ax^2}{1 + ax^2}\right)}. \quad (55)$$

Taking the square and $\log(\cdot)$ on both sides of (55), we obtain:

$$ax^4 + \left(\frac{4}{\pi} + a\log\left(1 - \left(P_{NN}(T)\right)^{\frac{2}{M}}\right)\right)x^2 + \log\left(1 - \left(P_{NN}(T)\right)^{\frac{2}{M}}\right) = 0.$$

The previous equation can be solved by introducing the substitution $t = x^2$. There is only one positive solution (out of four) which represents the threshold value:

$$T = \sigma_N\sqrt{\left(-4/\pi - aL + \sqrt{(4/\pi + aL)^2 - 4aL}\right)/a}, \quad (56)$$

which is an approximation of the threshold (53) with $L = \log\left(1 - \left(P_{NN}(T)\right)^{2/M}\right)$ and $a \approx 0.147$ suitable for hardware realizations.

### B. The single-pass threshold-based reconstruction procedure

The previous analysis can be used to define a simple CS reconstruction procedure. The threshold $T$ is used to determine the positions $\mathbf{P} = \{p_1, p_2, ..., p_K\}$ of signal components in the Hermite transform domain. If the estimated set of positions is such that $\mathbf{P} \subseteq \hat{\mathbf{P}}$ and $\text{card}\{\hat{\mathbf{P}}\} \leq M_A$ with $K \ll M$, the reconstruction can be achieved using the pseudo-inversion (11). The reconstruction procedure is presented with the following pseudo-code:

**Input:**
Signal length $M$, number of available samples $M_A$, transform matrix $\mathbf{\Psi}$, available samples positions $\mathbf{M_A} = \{m_1, m_2, ..., m_{M_A}\}$, measurement vector $\mathbf{y}_{cs}$.

Measurement matrix is: $\mathbf{A}_{cs} = \mathbf{\Psi}(\mathbf{M_A})_r$, where $(\cdot)_r$ denotes that only rows $\mathbf{M_A}$ are used from $\mathbf{\Psi}$.

**Output:**
1. Set $P_{NN}(T) \leftarrow 0.99$
2. $\mathbf{c}_0 \leftarrow \mathbf{A}_{cs}^{-1}\mathbf{y}_{cs}$

3. $\sigma_N \leftarrow \sqrt{\dfrac{M_A M - M_A^2}{M^2(M-1)} \sum_{p=0}^{M-1} \dfrac{M}{M_A} |c_p|^2}, c_p \in \mathbf{c}_0$

4. **Set** $a \leftarrow 0.147$

5. $L \leftarrow \log\left(1 - (P_{NN}(T))^{2/M}\right)$

6. $T \leftarrow \sigma_N \sqrt{\left(-4/\pi - aL + \sqrt{(4/\pi + aL)^2 - 4aL}\right)/a}$

7. $\hat{\mathbf{p}} \leftarrow \arg\{|\mathbf{c}_0| > T\}$

8. $\mathbf{A}_{csK} \leftarrow \mathbf{A}_{cs}(\hat{\mathbf{p}})_k$, only columns with indexes $\hat{\mathbf{p}}$ are used

9. $\mathbf{c}_K \leftarrow \left(\mathbf{A}_{csK}^T \mathbf{A}_{csK}\right)^{-1} \mathbf{A}_{csK}^T \mathbf{y}_{cs}$

**return** $\mathbf{c}_K$, $\hat{\mathbf{p}}$.

The reconstructed coefficient vector contains values $\mathbf{c}_K$ at positions $\hat{\mathbf{p}}$, and zeros at other positions.

## V. EXAMPLES

In order to validate the accuracy of variances derived in Section 3, statistical analysis was performed with respect to the number of available samples, order of Hermite coefficients and signal length. Probabilities of detection error and their respective approximations are verified by experimental examples. The performance of reconstruction algorithm based on the derived threshold is demonstrated in the last example.

*Example 1*: Consider the case of mono-component signal with unit amplitude that is sparse in the Hermite transform domain:

$$s(m) = \psi_{p_0}(m). \quad (57)$$

The signal has $M_A$ out of $M$ available samples. The Hermite function order $p_0$ is changed between:

(a) 0 and 199 for signal of length $M = 200$;
(b) 0 and 399 for signal of length $M = 400$.

For every given value $p_0$, 7000 independent realizations of the signal are performed, with $M_A$ available samples at random positions different in each realization, and the experimental value of the variance $\overline{\sigma}_s^2$ at the position $p_0$ is calculated.

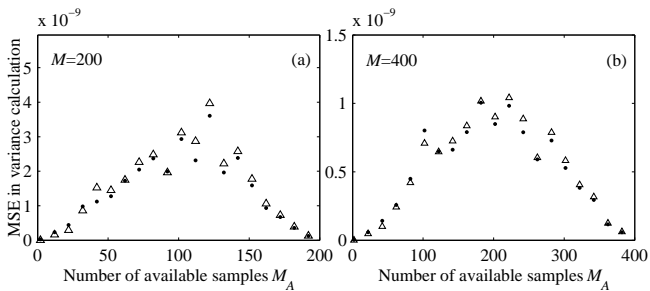

Fig. 4: The MSE of the variance calculation using (37) and (38) for the Hermite coefficient at the signal position $p_0$, for the different number of available samples. For given $M_A$ MSE is calculated for all possible signal coefficient positions $p_0$. Signals of length (a) $M$=200 and (b) $M$=400 are considered. For every position $p_0$ numerical variance values are calculated based on 7000 different realizations with randomly positioned samples.

The experimentally obtained variance is compared with the theoretical variance $\sigma_s^2$ given by (37) and its approximation (estimated value from available samples averaged over 7000 realizations) given by (38), based on the MSE calculation. The results are shown on Fig. 4a and Fig. 4b. The comparison is performed for different numbers of available samples: (a) between 2 and 200 with step 2, and (b) between 4 and 400 with step 4. Dotted line represents the MSE between the experimental results and theoretical model (37), while triangle line represents the MSE between experimental results and approximate model (38) with the assumption of known $p_0$. It can be seen that for both cases, the achieved MSE is of order $10^{-9}$, which confirms the accuracy of the derived theoretical variances.

*Example 2*: The mono-component signal of the form (57) is considered, for three Hermite coefficient positions (a) $p_0 = 1$, (b) $p_0 = 266$ and (c) $p_0 = 390$. The signal length is $M = 400$. The number of available samples $M_A$ is changed between 1 and $M$. For every given $M_A$, the variance of random variable $Y_{p=p_0}$ is calculated experimentally based on 5000 independent realizations of signal, with random missing samples positions. The variance $\tilde{\sigma}_s^2$ corresponding to signal position is calculated by (38) for every realization of signal. The results are averaged over 5000 realizations, for every given $M_A$. The results for the variances of $Y_{p=p_0}$ are presented in Fig. 5, for (a) $p_0 = 1$, (b) $p_0 = 266$ and (c) $p_0 = 390$. A significant matching of theoretical and experimental results is achieved for all signal positions $p_0$. Moreover, the variance of the non-signal coefficients is also statistically evaluated for all three mono-components signals, based on the same signal realizations. Experimental results along with theoretical variance $\sigma_N^2$ (given by (27)) are shown in Fig. 5d, confirming the fact that the variance $\sigma_N^2$ is independent of $p_0$.

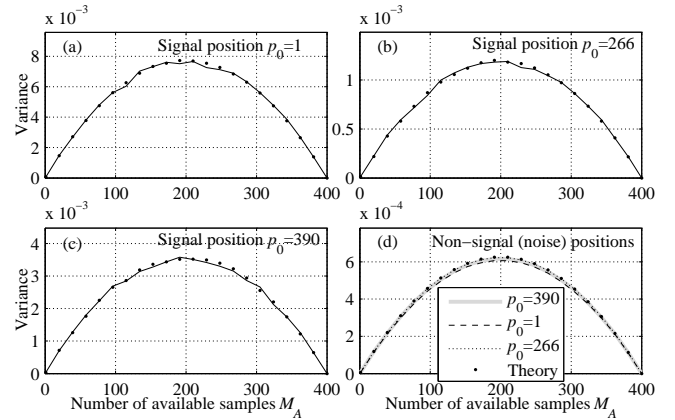

Fig. 5: The variance of the Hermite coefficient at signal component position as a function of available samples $M_A$; Different signal component positions are considered (a)-(c); Numerical result is denoted by black line, theoretical variance (38) is denoted by dots; (d) shows the numerically obtained variance of the non-signal Hermite coefficients $p \neq p_0$ for $p_0$ as in: (a) - denoted with dashed line, (b) - dotted line and (c) - tick gray line, as well as the theoretically calculated variance (27) of non-signal coefficients (dots).

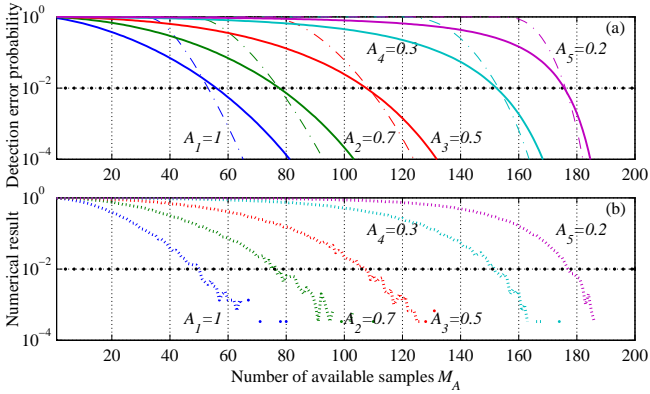

Fig. 6: Probability of components misdetection, presented as a function of the number of available samples: (a) exact probability calculated by (51) (solid line) and approximation (52); (b) experimental results

***Example 3***: The derived statistical parameters along with the error probability (51) and its approximation (52) are verified experimentally in this example. The signal with $M = 200$ samples and $K = 5$ components is given by:

$$s(m) = \sum_{i=1}^{K} A_i \psi_{p_i}(m) \qquad (58)$$

with $A_i=\{1, 0.7, 0.5, 0.3, 0.2\}$ and $p_i = \{20, 54, 94, 162, 192\}$ for $i = 1,…,K$. Fig. 6a shows the probability of misdetection for each component separately, calculated using (51). The probability approximation (52) is calculated as well. The number of available samples $M_A$ is varied between 1 and 200. Horizontal dotted line denotes the error probability equal to $P=10^{-2}$. It can be concluded that the exact and approximate probabilities almost match for the given probability $P=10^{-2}$. Note that, Fig. 6a specifies the number of available samples needed for successful detection of observed signal component with given probability. For example, 80 available samples are sufficient to detect the component with amplitude $A_1 = 0.1$ error probability close to 0, and the component with amplitude $A_2 = 0.7$ with error probability equal to $P=10^{-2}$. We can also conclude that about 176 available samples are needed for detection of all signal components with a given probability.

The probabilities are further experimentally evaluated. For every number of available samples $M_A$ between 1 and 200, the randomly positioned available signal samples were selected in 3000 realizations. In every realization, and for every signal component, the component misdetection events are counted. The misdetection of the $i$-th signal component occurs if at least one non-signal Hermite coefficient at position $p \neq p_i, i = 1,...,5$ has equal or higher amplitude than the the amplitude of the $i$-th signal component at $p = p_i$. The number of misdetection events is then divided by the number of signal realizations. The experiment is repeated for every $M_A$. Results are shown on Fig. 6b. Note that numerically obtained results almost match the theoretical ones in Fig. 6a.

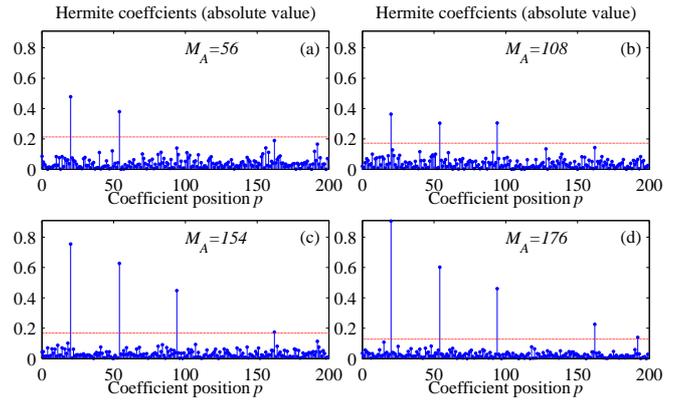

Fig. 7: Illustration of the automated threshold setting based on the number of available samples $M_A$

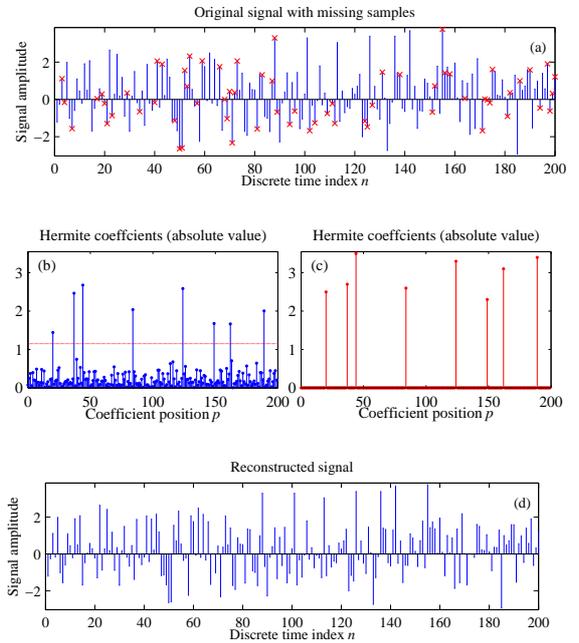

Fig. 8: (a) original signal with missing samples denoted with crosses, (b) Hermite coefficients of the signal with missing samples and threshold, (c) reconstructed transform and (d) reconstructed signal.

***Example 4***: Considered is the signal with missing samples from *Example 3*. Observed are different numbers of available samples used to calculate the expected probabilities of detection error (Fig. 6).

The first considered case is (a) $M_A = 56$ which enables detection of the signal components with the following probabilities of detection error: $P_1 = 0$, $P_2 = 0.0086$, $P_3 = 0.8679$, $P_4 = 1$ and $P_5 = 1$, for 5 considered signal components. This means that the 1st and the 2nd component will be detected with probability higher than 0.99, the 3rd component will be detected with probability ~ 0.13, while the 4th and the 5th component almost certainly will not be detected.

Similar discussion holds for:

(b) $M_A = 108$ where probabilities of detection error for different components are: $P_1 = 0$, $P_2 = 0$, $P_3 = 0.0109$, $P_4 = 1$ and $P_5 = 1$;

(c) $M_A = 154$ with detection error probabilities $P_1 = 0$, $P_2 = 0$, $P_3 = 0$, $P_4 = 0.0073$ and $P_5 = 0.9944$;

(d) $M_A = 176$, with corresponding detection error probabilities $P_1 = 0$, $P_2 = 0$, $P_3 = 0$, $P_4 = 0$ and $P_5 = 0.0106$. In this case, in about 99% of signal realizations all signal components will be above the threshold. The Hermite transform coefficients and probabilistic threshold (56) are shown on Fig. 7a-d for single signal realizations with different number of available samples.

*Example 5*. Our experimental analysis will be concluded through an example demonstrating the efficiency of the signal reconstruction algorithm. Consider the signal of the total length $M = 200$, having the form (58) with $A_i = \{2.5, 3.3, 2.6, 3.1, 2.7, 3.5, 2.3, 3.4\}$ and $p_i = \{20, 124, 84, 162, 37, 44, 149, 189\}$ for $i = 1,…,K$, and $K = 8$, $M_A = 135$ (32.5 % missing samples at random positions). Based on approximation (52) the given number of missing samples is sufficient for successful detection of all signal components with error probability lower than $10^{-2}$. Reconstruction results are shown in Fig. 8. Note that the reconstruction MSE is of order $10^{-24}$.

## VI. CONCLUSION

The paper analyzes the influence of missing samples of the compressed sensed signal to the Hermite domain representation. The effects of compressed sensing are statistically modelled in the Hermite sparsity domain using two independent random variables located at the signal and non-signal positions. Being able to characterize these variables allows us to develop a method to distinguish between them, and consequently, to easily determine the true signal support in the transform domain. Also, it was shown that depending on the percent of available samples and signal component amplitudes, we can calculate the probability of exact signal support detection. Furthermore, a very simple method for signal reconstruction is proposed based on the derived theoretical concepts. The crucial segments of presented theory are verified using a large number of statistical tests. Also, the efficiency of the proposed algorithm is proved on the examples.

*Acknowledgment*

This work is supported by the Montenegrin Ministry of Science, project grant: "New ICT Compressive sensing based trends applied to: multimedia, biomedicine and communications".